\documentclass[symmetry,review,accept,moreauthors,10pt,a4paper,aas_macros,symmetry]{Definitions/mdpi}
\usepackage{amssymb,color,graphicx,bm,mathrsfs,color}
\newcommand{\be}{\begin{equation}}
\newcommand{\ee}{\end{equation}}
\newcommand{\bea}{\begin{eqnarray}}
\newcommand{\eea}{\end{eqnarray}}

\newcommand{\ie}{{i.e.}}

\def\apj{ApJ~}%
\def\prc{Phys.~Rev.~C~}%
\def\prd{Phys.~Rev.~D~}%
\def\prl{Phys.~Rev.~Lett.}%
\usepackage[normalem]{ulem}  
\firstpage{494} 
\pubvolume{15}
\issuenum{2}
\articlenumber{494}
\pubyear{2023}
\copyrightyear{2023}
\externaleditor{Academic Editor: {Iver H. Brevik} 
}
\datereceived{27 December 2022} 
\daterevised{3 February 2023} 
\dateaccepted{9 February 2023} 
\datepublished{13 February 2023} 
\hreflink{https://doi.org/10.3390/sym15020494} 

\Title{Phenomenological Relativistic Second-Order Hydrodynamics for
       Multiflavor Fluids}
\TitleCitation{Phenomenological Relativistic Second-Order Hydrodynamics for
       Multiflavor Fluids}

\Author{Arus 
 Harutyunyan $^{1,2,}$*\orcidA{} and Armen Sedrakian $^{3,4}$\orcidB{}}

\AuthorNames{Arus Harutyunyan and Armen Sedrakian}
\AuthorCitation{Harutyunyan, A.; Sedrakian, A.}
\address{%
$^{1}$ \quad Byurakan Astrophysical Observatory, National Academy of
Sciences, Byurakan 0213, Armenia
\\
$^{2}$ \quad Department of Physics, Yerevan State University, 
  Yerevan 0025, Armenia\\
$^{3}$ \quad Frankfurt Institute for Advanced Studies, D-60438
  Frankfurt am Main, Germany 
\\
$^{4}$ \quad Institute of Theoretical Physics, University of Wroc\l{}aw,
50-204 Wroc\l{}aw, Poland; sedrakian@fias.uni-frankfurt.de, armen.sedrakian@uwr.edu.pl}
\corres{Correspondence: arus@bao.sci.am }

\abstract{ In this work, we perform a phenomenological derivation of
  the first- and second-order relativistic hydrodynamics of dissipative
  fluids. To set the stage, we start with a review of the ideal
    relativistic hydrodynamics from energy--momentum and particle
    number conservation equations. We then go on to discuss the
    matching conditions to local thermodynamical equilibrium,
    symmetries of the energy--momentum tensor, decomposition of
    dissipative processes according to their Lorentz structure, and,
    finally, the definition of the fluid velocity in the Landau and
    Eckart frames.  With this preparatory work, we first formulate the
    first-order (Navier--Stokes) relativistic hydrodynamics from
    the entropy flow equation, keeping only the first-order gradients of
    thermodynamical forces. A generalized form of diffusion terms is
    found with a matrix of diffusion coefficients describing the
    relative diffusion between various flavors.  The procedure of
    finding the dissipative terms is then extended to the second order to
    obtain the most general form of dissipative function for
    multiflavor systems up to the second order in 
    dissipative fluxes. The dissipative function now includes in
    addition to the usual second-order transport coefficients of
    Israel--Stewart theory also second-order diffusion between
    different flavors. The relaxation-type equations of second-order
    hydrodynamics are found from the requirement of positivity of the
    dissipation function, which features the finite relaxation times of
    various dissipative processes that guarantee the causality and
    stability of the fluid dynamics. These equations contain a
    complete set of nonlinear terms in the thermodynamic gradients and dissipative
    fluxes arising from the entropy current, which are not present in
    the conventional Israel--Stewart theory.
}

\begin{document}

\maketitle

\section{Introduction}
\label{sec:hydro_intro}

Relativistic hydrodynamics has been widely applied in recent years in
heavy-ion
physics~\cite{Jaiswal:2016hex,Florkowski_2018,Denicol2021,Elfner:2022iae}
and astrophysics~\cite{Font:2000pp,Rezzolla2013,Baiotti2017RPPh}.  For
example, the experiments at Relativistic Heavy Ion Collider (RHIC) and
Large Hadron Collider (LHC) motivated significantly the development of
the relativistic hydrodynamics with the data being well-described in
terms of a fluid with low shear viscosity; for reviews,
see~\cite{Busza_2018,Heinz2015IJMPE}.  Relativistic hydrodynamics is
also used in the general relativistic simulations of compact stars in
isolation or binaries, particularly in the context of gravitational
wave emission by neutron star mergers observed in
2017~\cite{Baiotti2019PrPNP}. A further astrophysical area of
applications of relativistic hydrodynamics is the physics of compact
star rotational dynamics, specifically glitches and their relaxations;
for reviews,~see~\cite{Haskell:2017lkl,Andersson:2020phh}.

The hydrodynamic description of fluids is valid close to the local thermal
equilibrium. The hydrodynamic state of a relativistic fluid is
described through its energy--momentum tensor and currents of
conserved charges, which in the low-frequency and long-wavelength
can be Taylor-expanded around their equilibrium values in
thermodynamic gradients (so-called thermodynamic forces). The validity
of such gradient expansion is guaranteed due to the clear separation
between the typical microscopic and macroscopic scales of the
system. The zeroth-order term in this expansion corresponds to the
limit of the ideal~hydrodynamics.

\textls[-15]{The truncation of the gradient expansion at the first order leads to
the relativistic Navier--Stokes (NS) theory, which was worked out by
Eckart~\cite{1940PhRv...58..919E} and
Landau--Lifshitz~\cite{Landau1987}.  It is known that the solutions of
the relativistic NS equations are acausal and
unstable}~\cite{1983AnPhy.151..466H,1985PhRvD..31..725H,1987PhRvD..35.3723H,Denicol2008JPhG}.
The reason for the acausality is the parabolic structure of NS
equations, which originated from the linear constitutive relations
between the dissipative fluxes and the thermodynamic forces.  Recent
work demonstrated that the acausalities and instabilities in
relativistic hydrodynamics are a consequence of the matching procedure
to the local equilibrium reference state.  More general matching
conditions were used to render the theory causal and stable at
first order~\cite{Bemfica:2019knx,Kovtun:2019hdm,Noronha2022PhRvL}.

The problem of acausality can be solved in the second-order theory,
where additional terms appear that contain higher (second)-order
derivatives in thermodynamic quantities. For non-relativistic fluids,
the second-order theory was proposed by
M\"uller~\cite{1967ZPhy..198..329M}, and then rediscovered and
extended to relativistic systems by Israel and
Stewart~\cite{1976AnPhy.100..310I,1979AnPhy.118..341I}. In these
theories, the dissipative fluxes are treated as independent state
variables that satisfy relaxation-type equations derivable from the
entropy principle. The relaxation terms which appear in these
equations recover the causality of the
theory~\cite{Denicol2008JPhG,Pu2010PhRvD}.  The relaxation equations
for dissipative fluxes in the second-order theories and their numerical
studies in simulations have been discussed extensively in the
literature; see~\cite{Baier2008JHEP,Betz2009JPhG,Romatschke2010CQGra,Tsumura2010PhLB,Betz2011EPJWC,Moore2011PhRvL,Moore2012JHEP,Jaiswal2013PhRvC,Jaiswal2015PhLB,Florkowski2015PhRvC,Finazzo2015JHEP,Tinti2017PhRvD}.

The relativistic second-order hydrodynamics can be obtained from
moments of the Boltzmann equation for the distribution
function~\cite{Denicol2012PhRvD,Denicol2012PhLB,Molnar2016PhRvD}.
This theory provides a systematic way of evaluating the new
coefficients describing the relaxation of dissipative quantities at
weak coupling and in the quasiparticle limit.  An alternative
approach valid in the strong coupling limit is Zubarev's
non-equilibrium statistical operator formalism, which was recently
applied to obtain the second-order hydrodynamics and/or Kubo-type
formulae for transport
coefficients~\cite{Harutyunyan:2018cmm,Becattini2019,Harutyunyan:2021rmb}.
Related approaches based on field theory methods are given in
Refs.~\cite{Tokarchuk2018,Calzetta2022}.

{ Recent applications of relativistic hydrodynamics focused on the
  systems featuring spin, polarization, and vorticity. A systematic
  calculation of the corrections of the stress--energy tensor and
  currents up to the second order in thermal vorticity were given in
  Refs.~\cite{Buzzegoli2017JHEP,Becattini2021} and the relevant Kubo
  formulae were derived. Similarly, the problem of relativistic
  hydrodynamics under strong spin was studied on the basis of
  entropy--current analysis in Ref.~\cite{Cao2022PTEP} where the
  seventeen transport coefficients of highly anisotropic relativistic
  hydrodynamics were identified. Ref.~\cite{Hongo2021JHEP} formulated
  relativistic spin hydrodynamics of Dirac fermions in a torsionful
  curved background and derived the relevant Kubo formulae associated
  with the correlation functions. The spin hydrodynamics was also
  derived from the Boltzmann equation using the method of moments in
  Ref.~\cite{Weickgenannt2022} up to the second order. Plasmas with
  vorticity were studied in second-order dissipative hydrodynamics in
  the presence of a chiral imbalance in Ref.~\cite{Gorbar2017PhRvD}
  and the dispersion relations of chiral vortical waves were
  obtained.}

{Specific applications of second-order hydrodynamics include,
  for example, the computation of the transport coefficients from a
  Chapman--Enskog-like expansion for a system of massless quarks and
  gluons~\cite{Jaiswal2015PhLB}. The second-order hydrodynamics was
  recently applied to multicomponent systems with hard-sphere
  interactions in Ref.~\cite{Hu2022PhRvD}, where the multicomponent
  nature of the system was taken into account. The second-order
  hydrodynamics was applied to quark--gluon plasma in heavy ion
  collisions, for example, in Ref.~\cite{Almaalol2022}, showing the
  coupling between diffusion and shear and bulk viscosities in 
  a multi-component system. A discussion of the use of relativistic anisotropic
  hydrodynamics to study the physics of ultrarelativistic heavy-ion
  collisions was given in Ref.~\cite{Alalawi2022}, using as 
  the main ingredient the quasiparticle anisotropic hydrodynamics
  model for quark--gluon plasma.  Anomalies in the field theory were
  included in the description of heavy-ion collisions in
  Ref.~\cite{Buzzegoli2022}.  The strong gravity regime of second-order
  hydrodynamics, relevant for astrophysical applications, was studied
  in Ref.~\cite{Lahiri2020}. The first-order theory was applied in the
  context of the out-of-equilibrium dynamics of viscous fluids in a
  spatially flat cosmology~\cite{BemficaPhysRevD}.  }

{ Recent work also extended the current formulations of
  relativistic hydrodynamics to include the effect of magnetic
  fields. The first-order dissipative effects and relevant transport
  coefficients were derived, and a numerical method was developed using
  the method of moments in Ref.~\cite{MostPhysRevD}. The effects of anomalies
  in the magnetohydrodynamics were included in
  Ref.~\cite{Landry2022}. Charge diffusion in the second-order
  theories was recently studied in
  Ref.~\cite{Dash2022}. Applications to neutron stars and relevant
  transport coefficients were derived, for example, in
  Refs.~\cite{Harutyunyan2016PhRvC,Harutyunyan2015,Shternin2022EPJA}.
  A comprehensive review of the new developments in this field is
  given in Ref.~\cite{Hattori2022S}. }

More formal recent developments of relativistic hydrodynamics
  include formulations that avoid specific frames. (The choice of
  frames in relativistic heavy ion collisions was discussed in
  Ref.~\cite{Monnai2019}.) First- and second-order viscous relativistic
  hydrodynamics were formulated without specific frame conditions in
  Refs.~\cite{BemficaPhysRevX,Noronha2022PhRvL}, respectively,
  with the causality and stability conditions explicitly formulated in
  the conformal regime in the second work.  Earlier, the stability of
  first-order hydrodynamics was established in
  Ref.~\cite{Kovtun:2019hdm}.
  
The purpose of this article is twofold. First, we review the
phenomenological derivation of the second-order dissipative
hydrodynamics from general principles of conservation laws and the
second law of thermodynamics. The dissipative equations are obtained
from expansions of the energy--momentum tensor and currents around
their equilibrium values in thermodynamic gradients (so-called
thermodynamic forces) order by order. Second, we consider the theory
in the general case of a system with $l$ independent flavors of
conserved charges, { keeping all possible second-order terms which
  arise from the entropy current and the second law of thermodynamics.
  In the case where the nonlinear terms in the thermodynamic gradients
  and dissipative fluxes are dropped, these equations are reduced to
  the original Israel and Stewart
  hydrodynamics~\cite{1976AnPhy.100..310I,1979AnPhy.118..341I}.}

This work is organized as follows. In Section~\ref{sec:ideal_hd}, we
present the ideal hydrodynamics of non-dissipative fluids. Dissipation
in the fluids is introduced in Section~\ref{sec:diss_hydro}. We discuss
the first-order hydrodynamics and obtain the Navier--Stokes equations
in Section~\ref{sec:NS}.  The second-order Israel--Stewart theory in the
case of a system with $l$ independent flavors of
conserved charges is given in Section~\ref{sec:IS}. A summary is provided
in Section~\ref{sec:summary}.

\section{Relativistic Ideal Hydrodynamics}
\label{sec:ideal_hd}

Relativistic hydrodynamics describes the state of the fluid
employing its energy--momentum tensor $T^{\mu\nu}$ and currents 
of conserved charges $N^{\mu}_a$, such as the baryonic, electric, 
etc. Here, we consider the general case of a system 
with $l$ independent flavors of conserved charges, which are 
labeled by the index $a$. The equations of relativistic 
hydrodynamics are contained in the conservation laws for 
the energy--momentum tensor and the charge~currents 
\be\label{eq:T_Na_cons}
\partial_{\mu} T^{\mu\nu} =0,\qquad
\partial_{\mu} N^{\mu}_a=0,\quad a=1,2,...,l.
\ee
In dissipative hydrodynamics, also the entropy principle $\partial_\mu
S^\mu\geq 0$ should be applied to close the
system~\eqref{eq:T_Na_cons}, where $S^\mu$ is the entropy 4-current.

Ideal hydrodynamics corresponds to the zeroth-order expansion of the
energy-\linebreak momentum tensor and charge currents with respect to the
thermodynamic forces. In this case, each fluid element maintains the
local thermal equilibrium during its
evolution~\cite{Landau1987,1972gcpa.book.....W}. The macroscopic state
of the fluid is therefore fully described through the fields of
the energy density $\epsilon(x)$ and the charge densities $n_a(x)$. 
The fluid 4-velocity is defined as
\bea\label{eq:4_vel}
u^\mu(x)=\frac{d x^\mu}{d\tau}=\left(\gamma(x),\gamma v^i(x)\right),
\eea
where $x^\mu$ is the displacement 4-vector of a fluid element,
$d\tau=\sqrt{g_{\mu\nu}dx^\mu dx^\nu}$ is the infinitesimal change in
the proper time, $\bm v(x)$ is the fluid 3-velocity and
$\gamma(x)=(1-\bm v^2)^{-1/2}$ is the relevant Lorentz factor. The
4-velocity given by Equation~\eqref{eq:4_vel} is normalized by the
condition $u^\mu(x) u_\mu(x)=1$ and has only three independent
components.

Because the thermal equilibrium is maintained locally, each fluid
element can be assigned well-defined local values of the temperature
$T(x)$, chemical potentials $\mu_a(x)$ (conjugate to charge densities
$n_a(x)$), entropy density $s(x)$ and pressure $p(x)$. These
quantities are related to the local energy and charge densities via an
equation of state $p=p(\epsilon, n_a)$ and the standard thermodynamic
relations
\bea\label{eq:thermodyn1}
Tds &=& d\epsilon -\sum\limits_a\mu_a dn_a,\hspace{1.105cm} ({\rm first~law})\\
\label{eq:thermodyn2}
dp &=& sdT+\sum\limits_an_a d\mu_a,\qquad ({\rm Gibbs-Duhem~relation})\\
\label{eq:thermodyn3}
h &=& \epsilon +p=Ts+\sum\limits_a\mu_a n_a,
\eea
where $h$ is the enthalpy density. 

The fluid 4-velocity is defined in such a way that in the fluid rest
frame, the energy and charge flows vanish: $N^i=0$ and $T^{0i}=0$, if
$u^i=0$. These conditions together with the spatial isotropy imply the
following form for the energy--momentum tensor and charge~currents~\cite{Landau1987,1972gcpa.book.....W}:
\bea \label{eq:T_N_0}
{T}^{\mu\nu}_0 = \epsilon u^{\mu}u^{\nu} - p\Delta^{\mu\nu},\qquad
{N}^{\mu}_{a0} = {n}_au^\mu,
\eea
where the index $0$ labels the quantities in ideal hydrodynamics;
$\Delta^{\mu\nu}=g^{\mu\nu}-u^\mu u^\nu$ is the projection operator
onto the 3-space orthogonal to $u_\mu$ and has the properties
\bea\label{eq:prop_proj}
u_\mu\Delta^{\mu\nu}=\Delta^{\mu\nu}u_\nu=0,\qquad
\Delta^{\mu\nu}\Delta_{\nu\lambda}=\Delta^{\mu}_\lambda,\qquad \Delta^{\mu}_{\mu}=3.
\eea 
From Equations~\eqref{eq:T_N_0} and \eqref{eq:prop_proj}, we obtain
\bea  \label{eq:ideal_therm}
\epsilon = u_\mu u_\nu{T}^{\mu\nu}_0,\qquad
p =-\frac{1}{3}\Delta_{\mu\nu}
{T}^{\mu\nu}_0,\qquad
{n}_a = u_\mu{N}^{\mu}_{a0}.
\eea
In the fluid rest frame, $u^\mu=(1,0,0,0)$; therefore,
$\Delta^{\mu\nu}={\rm diag}(0,-1,-1,-1)$ and
$\Delta^{ij}=\Delta_{ij}=-\delta_{ij}$. In this case,
Equation~\eqref{eq:ideal_therm} simplifies to
\bea  \label{eq:ideal_therm_rest}
\epsilon = {T}^{00}_0,\qquad
p =\frac{1}{3}{T}^{ii}_0,\qquad
{n}_a = {N}^{0}_{a0}.
\eea
We introduce also the energy 4-current or momentum density by the formula
\bea\label{eq:momentum_dens} P^\mu_0 =u_\nu T^{\mu\nu}_0+pu^\mu =h
u^\mu, \eea
which is the relativistic generalization of the 3-momentum density
(mass current).  As seen from Equations~\eqref{eq:T_N_0} and
\eqref{eq:momentum_dens}, all charge currents are parallel to each
other and to the energy flow, which is due to the possibility of a
unique definition of the velocity field for ideal~fluids.
 
The equations of ideal hydrodynamics are obtained by substituting the
expressions \eqref{eq:T_N_0} into the conservation
laws~\eqref{eq:T_Na_cons}
\bea \label{eq:ideal_hydro1}
\partial_{\mu}{N}^{\mu}_{a0}= \partial_{\mu} (n_au^\mu)=0,\qquad
\partial_{\mu}{T}^{\mu\nu}_0=
\partial_{\mu}(hu^{\mu}u^{\nu}-pg^{\mu\nu}) = 0.
\eea

Contracting the second equation in \eqref{eq:ideal_hydro1} once with
$u_\nu$ and then with the projector $\Delta_{\alpha\nu}$ and taking
into account Equation~\eqref{eq:prop_proj} and the condition
$u_\nu\partial_\mu u^{\nu}=0$, we obtain
\bea \label{eq:ideal_hydro2}
Dn_a+ n_a\theta =0,\qquad
D\epsilon+ h \theta =0,\qquad
h D u_\alpha = \nabla_\alpha p,
\eea
where we introduce the covariant time derivative, covariant
spatial derivative, and velocity 4-divergence via $D=u^\mu\partial_\mu$, $\nabla_\alpha=
\Delta_{\alpha\beta}\partial^\beta$ and 
$\theta =\partial_\mu u^\mu$, respectively. The velocity 4-divergence
$\theta$ quantifies how fast the fluid is expanding ($\theta > 0$)
or contracting ($\theta < 0$); it is called the {\it fluid expansion rate}.
In the case of an incompressible flow of the fluid, we have $\theta=0$.

It is not difficult to recognize in the first two relations in
\eqref{eq:ideal_hydro2} the covariant expressions for the charge
conservation law and the energy conservation law, respectively. The
third equation is nothing more than the relativistic generalization of
the ordinary Euler equation familiar from nonrelativistic
hydrodynamics. From these equations, one can deduce that in
relativistic hydrodynamics, the role of rest mass density is taken
over by the enthalpy density $h$, which thus provides the correct
inertia measure for relativistic fluids. This fact illustrates the
importance of the quantity \eqref{eq:momentum_dens} as the
relativistic analog of the momentum~flux.

The system \eqref{eq:ideal_hydro2} contains $l+4$ equations for $l+5$
variables $\epsilon$, $p$, $n_a$ and $u^\mu$. To close the system, one
still needs to specify an equation of state $p=p(\epsilon, n_a)$, which
relates the pressure to the conserved thermodynamic variables.

It is easy to show that the equations of ideal hydrodynamics lead
automatically to entropy conservation.  The entropy flux can be
written as
\bea\label{eq:s_mu_ideal}
S^\mu_0 =su^\mu.
\eea
Using the thermodynamic relations
\eqref{eq:thermodyn1}--\eqref{eq:thermodyn3} and the equations of
motion~\eqref{eq:ideal_hydro2}, we obtain
\bea\label{eq:ent_gen_ideal}
T\partial_\mu S^\mu_0 =Ds+s\theta =
 D\epsilon +h\theta
 - \sum\limits_a\mu_a(n_a\theta +Dn_a) =0,\quad
\eea
\textls[-15]{which is the second law of thermodynamics for a non-dissipative
system.  Using} Equations~\eqref{eq:thermodyn3}, \eqref{eq:T_N_0},
\eqref{eq:momentum_dens} and \eqref{eq:s_mu_ideal}, we can rewrite the
entropy current as
\bea\label{eq:s_mu_ideal1}
S^\mu_0 =\beta P^\mu_0 -\sum\limits_a\alpha_a N^\mu_{a0}
=p\beta^\mu+\beta_\nu T^{\mu\nu}_0-\sum\limits_a\alpha_a N^\mu_{a0},
\eea
where we defined
\be\label{eq:beta_alpha}
\beta=T^{-1}, \qquad \beta^\nu= \beta u^{\nu},\qquad \alpha_a =\beta\mu_a.
\ee

\textls[-10]{The expression \eqref{eq:s_mu_ideal1} is the covariant form of the
relation \eqref{eq:thermodyn3}.  To proceed further, it is convenient
to modify Equations~\eqref{eq:thermodyn1} and \eqref{eq:thermodyn2} using
the definitions~\eqref{eq:beta_alpha}. We obtain} for
Equation~\eqref{eq:thermodyn2}
\bea
dp =-\beta^{-1}hd\beta
+\beta^{-1}\sum\limits_a n_ad\alpha_a,
\eea
where we used Equation~\eqref{eq:thermodyn3} in the second step.  Now the
first law of thermodynamics and the Gibbs--Duhem relation can be
written in an alternative form:
\bea\label{eq:thermodyn6}
ds=\beta d\epsilon -\sum\limits_a\alpha_a dn_a,\qquad
\beta dp  =-hd\beta +\sum\limits_a n_ad\alpha_a.
\eea

With the aid of Equations~\eqref{eq:T_N_0} and \eqref{eq:s_mu_ideal1},
 these relations can be cast into a covariant form:
\bea\label{eq:thermodyn1_cov}
dS^\mu_0 &=& \beta_\nu dT^{\mu\nu}_{0}-
\sum\limits_a\alpha_a d N^\mu_{a0},\\
\label{eq:thermodyn2_cov}
d(p\beta^\mu) &=& - T^{\mu\nu}_{0} d\beta_\nu +
\sum\limits_a N^\mu_{a0} d\alpha_a,
\eea
where we used the second relation in Equation~\eqref{eq:thermodyn6}.
One should note that, despite their vector form, these 
equations do not contain more information than the scalar 
thermodynamic relations (contraction of Equations~\eqref{eq:thermodyn1_cov} 
and \eqref{eq:thermodyn2_cov} with the projector $\Delta_{\mu\nu}$ 
leads to identities).

\section{Relativistic Dissipative Hydrodynamics}
\label{sec:diss_hydro}

\subsection{Matching Conditions}
\label{sec:matching}

Ideal hydrodynamics relies on the strong assumption of the local
thermodynamic equilibrium of each fluid element.  However, the local
equilibrium cannot be maintained permanently since all real systems
react to the non-uniformities of the fluid on finite time scales by
generating irreversible (dissipative) fluxes.  These fluxes tend to
eliminate the local gradients to drive the system toward
global thermal equilibrium, leading thereby to energy dissipation and
entropy increase.

As in the case of ideal hydrodynamics, the equations of relativistic
dissipative hydrodynamics can be obtained from the conservation laws
given by Equation~\eqref{eq:T_Na_cons}. The expressions~\eqref{eq:T_N_0}
then need to be generalized by taking into account all possible
effects of dissipation.  The energy--momentum tensor obtains an
anisotropic contribution because of irreversible momentum exchange
between different fluid elements, as well as heat transfer due to
the relative motion of energy and charge currents. The diffusion processes
between different charge species in their turn introduce dissipative
terms in the charge currents $N_a^\mu$. As a result, the
energy--momentum tensor and the charge currents for a dissipative fluid
can be written in the following form:
\bea \label{eq:T_munu_decomp1}
{T}^{\mu\nu} &=& {T}^{\mu\nu}_0 +\tau^{\mu\nu}={\epsilon} u^{\mu}u^{\nu} - p\Delta^{\mu\nu} + \tau^{\mu\nu},\\
\label{eq:N_a_decomp1}
{N}^{\mu}_a &=& {N}^{\mu}_{a0}+j_a^{\mu}
={n}_au^\mu +{j}^{\mu}_a,
\eea
where ${j}^\mu_a$ and ${\tau}^{\mu\nu}$ are dissipative terms, and the
tensor ${\tau}^{\mu\nu}$ is symmetric. Note that since the system is
out of thermodynamic equilibrium, the thermodynamic parameters are not
well-defined anymore, and one needs to impose additional conditions to
specify what should be exactly understood under $\epsilon$, $n_a$ and
$p$ in Equations~\eqref{eq:T_munu_decomp1} and \eqref{eq:N_a_decomp1}. The
thermodynamic variables in non-equilibrium states can be defined only
by means of a {\it fictitious} equilibrium state, which should be
constructed in such a way as to satisfy the thermodynamic
relations~\eqref{eq:thermodyn1}--\eqref{eq:thermodyn3}~\cite{1979AnPhy.118..341I}. In
order to construct such an equilibrium state for the given values of
${T}^{\mu\nu}$ and ${N}^{\mu}_a$, we define first the energy and
charge densities via the so-called matching (fitting) conditions:
\bea\label{eq:match_cond}
\epsilon(u)=u_\mu u_\nu T^{\mu\nu},\qquad
n_a(u)=u_\mu N^\mu_a.
\eea

Equations \eqref{eq:match_cond} imply simply that $\epsilon$ and $n_a$
are the time-like eigenvalues of the energy--momentum tensor and the
charge currents, respectively, measured by a local observer comoving
with the fluid element.  We remark that the quantities $\epsilon$ and
$n_a$ depend in general on the choice of $u^\mu$, which is the
consequence of the ambiguity of the definition of the velocity field
for dissipative fluids (see Section~\ref{sec:frames} for details).

We define in the next step an equilibrium entropy density via an
equation of state $s=s(\epsilon, n_a)$, which is chosen to be {\it the
  same function} of the parameters $\epsilon$ and $n_a$, as it would
be in full thermodynamic
equilibrium~\cite{1971ApJ...168..175W,1976AnPhy.100..310I,1979AnPhy.118..341I,1983AnPhy.151..466H}. The
rest of the local thermodynamic quantities can be defined through
standard thermodynamic
relations~\eqref{eq:thermodyn1}--\eqref{eq:thermodyn3}, \ie,
\bea\label{eq:non_eq_therm}
\beta = \left(\frac{\partial s}{\partial \epsilon}\right)_{n_a},\qquad
\mu_a = -T\left(\frac{\partial s}{\partial n_a}\right)_{\epsilon, n_b\neq n_a},\qquad
p = Ts+\sum\limits_a\mu_a n_a-\epsilon,
\eea
which implies that all thermodynamic parameters are defined in
non-equilibrium states via the {\it same functions} of $\epsilon$ and
$n_a$ as their equilibrium counterparts.

The thermodynamic parameters defined above form a fictitious
equilibrium state for which the {\it reversible} thermodynamic
relations are formally valid.  However, it is worthwhile to stress
that only the energy and charge densities can be ascribed a definite
physical meaning, whereas the quantities $s$, $p$, $T$, and $\mu_a$ do
not retain their usual physical meaning when the system is out of
thermal equilibrium.  These quantities are mathematically
convenient to use since they have the approximate physical meaning of
their equilibrium counterparts for small departures from the local
equilibrium~\cite{1976AnPhy.100..310I,1979AnPhy.118..341I}.

For instance, the quantity $p$ is not the {\it actual thermodynamic
  pressure} (\ie, the work done by a unit change of the fluid volume),
but differs from the latter by an additional non-equilibrium term,
which is of the first order in velocity gradients~\cite{Landau1987,
  1979AnPhy.118..341I}.  Similarly, the quantity $s$ is not the {\it
  actual non-equilibrium entropy density} since it is defined through
reversible thermodynamic relations. In other words, $s$ is not the
proper quantity which {\it should increase} in non-equilibrium
processes according to the second law of thermodynamics. However, the
first non-equilibrium correction, in this case, appears only at the
second order in gradients, and, therefore, can be ignored in the
first-order
theory~\cite{Landau1987,1976AnPhy.100..310I,1979AnPhy.118..341I}. Indeed,
the correction to $s$ should be negatively defined since the entropy
attains its maximum in equilibrium. The only scalar quantity which
might contribute to the entropy density at the first order in
gradients is the velocity 4-divergence $\theta$; therefore, the
non-equilibrium entropy density should be
$s'=s(\epsilon, n_a)-a\theta$, with some thermodynamic coefficient
$a$. Since $\theta$ can be of both signs, we conclude that $a= 0$.

\subsection{Decomposition in Different Dissipative Processes}
\label{sec:decomp}

To identify different dissipative processes, it is convenient
to separate scalar, vector, and traceless parts of the tensor
$\tau^{\mu\nu}$. We first note that the matching conditions
\eqref{eq:match_cond} together with Equations~\eqref{eq:T_munu_decomp1} and
\eqref{eq:N_a_decomp1} impose the following orthogonality conditions
on the dissipative terms:
\bea\label{eq:tau_orth}
u_\mu u_\nu\tau^{\mu\nu}=0,\qquad u_\mu j^\mu_a =0.
\eea

The tensor $\tau^{\mu\nu}$ can be further decomposed into its
irreducible components parallel and orthogonal to the fluid 4-velocity
$u^\mu$. For that purpose, it is useful to introduce a fourth-rank
traceless projector orthogonal to $u^\mu$ via
\bea\label{eq:projector_delta4}
\Delta_{\mu\nu\rho\sigma}= \frac{1}{2}\left(\Delta_{\mu\rho}\Delta_{\nu\sigma}
+\Delta_{\mu\sigma}\Delta_{\nu\rho}\right)
-\frac{1}{3}\Delta_{\mu\nu}\Delta_{\rho\sigma},
\eea
which has the properties 
\bea\label{eq:prop_projector4} 
\Delta_{\mu\nu\rho\sigma}=\Delta_{\nu\mu\rho\sigma}=\Delta_{\rho\sigma\mu\nu},\quad
u^\mu \Delta_{\mu\nu\rho\sigma}=0,\quad
\Delta_{\alpha}^{\mu} \Delta_{\mu\nu\rho\sigma}=
\Delta_{\alpha\nu\rho\sigma},\nonumber\\
\Delta_{\mu~\rho\sigma}^{~\mu}=0,\quad
\Delta_{\nu\mu~\sigma}^{\quad\mu}=
\frac{5}{3}\Delta_{\nu\sigma},\quad
\Delta_{\mu\nu}^{\quad\mu\nu}=5,\quad
\Delta_{\mu\nu\rho\sigma}\Delta^{\rho\sigma}_{\alpha\beta}
=\Delta_{\mu\nu\alpha\beta}.
\eea

The most general tensor decomposition of $\tau^{\mu\nu}$ consistent
with the orthogonality condition~\eqref{eq:tau_orth} can be now
written as
\bea \label{eq:tau_munu_decomp}
\tau^{\mu\nu} = -\Pi \Delta^{\mu\nu} 
+ q^{\mu}u^{\nu}+ q^{\nu}u^{\mu} + {\pi}^{\mu\nu},
\eea
where we defined new dissipative quantities via
\bea \label{eq:tau_munu_comp}
\Pi = -\frac{1}{3}\Delta_{\mu\nu}
\tau^{\mu\nu}=-\frac{1}{3}\tau^{\mu}_\mu, \qquad
q^\mu  = u_{\alpha}\Delta^\mu_{\beta}\tau^{\alpha\beta}, \qquad
\pi^{\mu\nu} = \Delta_{\alpha\beta}^{\mu\nu} 
\tau^{\alpha\beta},
\eea
and used the properties \eqref{eq:prop_proj}.  From
Equations~\eqref{eq:T_munu_decomp1}, \eqref{eq:N_a_decomp1} and
\eqref{eq:tau_munu_decomp}, we obtain the most general decompositions
of the energy--momentum tensor and the charge currents into their
irreducible components:
\bea \label{eq:T_munu_decomp2}
{T}^{\mu\nu} &=& {\epsilon} u^{\mu}u^{\nu} - ({p}+{\Pi})\Delta^{\mu\nu} 
+ q^{\mu}u^{\nu}+ q^{\nu}u^{\mu} + {\pi}^{\mu\nu},\\
\label{eq:N_a_decomp2}
{N}^{\mu}_a &=& {n}_au^\mu +{j}^{\mu}_a.
\eea
The energy flow defined in Equation~\eqref{eq:momentum_dens}
can be generalized for dissipative fluids as
\bea\label{eq:momentum_dens_diss}
P^\mu =u_\nu T^{\mu\nu}+pu^\mu =h u^\mu +q^\mu.
\eea

The dissipative terms ${j}^\mu_a$, $q^\mu$, ${\pi}^{\mu\nu}$ and
${\Pi}$ in Equations~\eqref{eq:T_munu_decomp2} and \eqref{eq:N_a_decomp2}
are called charge diffusion fluxes, energy diffusion flux, shear
stress tensor, and bulk viscous pressure, respectively. The shear
stress tensor is the traceless spatial part of the energy--momentum
tensor and describes the dissipation by anisotropic momentum flow,
whereas the bulk viscous pressure is the non-equilibrium part of the
pressure and is responsible for dissipation during isotropic expansion
or compression. The dissipative currents satisfy the following~conditions:
\bea \label{eq:orthogonality1}
u_{\nu}q^{\nu} = 0,\qquad u_{\nu}{j}^{\nu}_a = 0,\qquad 
u_{\nu}{\pi}^{\mu\nu} = 0;\qquad {\pi}_{\mu}^\mu=0,
\eea
the first three of which reflect the fact that the dissipation in the fluid should be spatial.

Finally, all quantities on the right-hand sides of
Equations~\eqref{eq:T_munu_decomp1} and \eqref{eq:N_a_decomp1} can be
obtained by the relevant projections of $T^{\mu\nu}$ and $N^\mu_a$:
\bea  \label{eq:proj1}
\epsilon = u_\mu u_\nu T^{\mu\nu},\qquad
n_a = u_\mu N^{\mu}_a,\qquad
p +\Pi =-\frac{1}{3}\Delta_{\mu\nu}
T^{\mu\nu},\\
\label{eq:proj2}
\pi^{\mu\nu} = \Delta_{\alpha\beta}^{\mu\nu} T^{\alpha\beta},\qquad
q^\mu  = u_\alpha\Delta_{\beta}^{\mu}T^{\alpha\beta},\qquad
j_a^{\nu}=\Delta_{\mu}^{\nu} N^{\mu}_a,\quad
\eea
as follows from Equations~\eqref{eq:T_munu_decomp1}--\eqref{eq:match_cond},
\eqref{eq:tau_munu_comp} and the properties \eqref{eq:prop_proj} and
\eqref{eq:prop_projector4}.  In the fluid rest frame, we have
\bea \label{eq:currents_rest1}
\epsilon = T^{00},\qquad
n_a = N^{0}_a,\qquad
p+\Pi =-\frac{1}{3}T^k_k,\hspace{1.cm}\\
\label{eq:currents_rest2}
\pi_{kl} = \left(\delta_{ki}\delta_{lj}-\frac{1}{3}\delta_{kl}\delta_{ij}
\right) T_{ij},\qquad
q^i  = T^{0i},\qquad
j_a^{i} = N^{i}_a.
\eea

From these expressions, it can be seen that all dissipative quantities
are, as expected, purely spatial in the rest state of the
fluid. Each of the vectors $q^\mu$ and $j_a^\mu$ thus has 3
independent components, and the shear stress tensor $\pi^{\mu\nu}$ has
5 independent components since its trace vanishes. Thus, the total
number of independent quantities in Equations~\eqref{eq:currents_rest1} and \eqref{eq:currents_rest2} is $4l+10$,
which corresponds to the degrees of freedom of the
energy--momentum tensor and the charge currents (remembering that $p$
is determined by an equation of state). As for the velocity of the
fluid, $u^\mu$ should not be treated as an independent variable but
should be associated with one of the physical currents. We devote the
next subsection to the discussion of the possible definitions of the
velocity field.

\subsection{Definition of Flow Velocity}
\label{sec:frames}

Another important question in dissipative fluid dynamics is the proper
definition of fluid velocity.  The choice of the frame in the case
of ideal hydrodynamics is simple, as it has the energy and charge
current flowing parallel to each other. Then, the fluid rest frame is
defined via the requirement that these currents vanish identically.
The situation is different in the case of dissipative fluids because
one is faced with simultaneous flows of energy and particle
currents. There are two simple and also natural ways to define the
fluid rest state in dissipative hydrodynamics, which we describe in
this subsection in turn.


In the \emph{Landau frame} (or L-frame)~\cite{Landau1987} the net energy flow
vanishes, the definition of fluid velocity $u^\mu$ in this frame is
then given through the time-like eigenvector of $T^{\mu\nu}$:
\bea\label{eq:vel_Landau}
u_L^\mu =\frac{u_{L\nu} T^{\mu\nu}}{\sqrt{u_{L\nu} T^{\mu\nu} u^{\lambda}_L T_{\mu\lambda}}},
\eea
which in combination with Equations~\eqref{eq:match_cond},
\eqref{eq:T_munu_decomp2}, and \eqref{eq:orthogonality1}, leads to the
following relations:
\bea\label{eq:eps_Landau}
\epsilon_L =\sqrt{u_{L\nu} T^{\mu\nu} u^\lambda_L T_{\mu\lambda}},\qquad
u_{L\nu} T^{\mu\nu}=\epsilon_L u_{L}^\mu,
\qquad q^\mu_L =0,
\eea
where the subscript $L$ indicates that the quantities are evaluated
according to the Landau definition of $u^\mu$.  With such a choice of
the velocity field, the energy diffusion flux is zero, whereas the heat
transport is accommodated in the particle diffusion fluxes
$j_{La}^\mu$.  A different form of Equation~\eqref{eq:vel_Landau} is
obtained by noting that the energy flux~\eqref{eq:momentum_dens_diss}
in the  L-frame is parallel to the flow velocity: 
\bea\label{eq:vel_Landau1}
u_L^\mu =\frac{P^{\mu}}{\sqrt{P^{\mu} P_{\mu}}}.
\eea

Next, consider the velocity of fluid $u^\mu$ in a generic frame. It can
be related to $u^\mu_L$ upon noticing that in a generic fluid rest
frame the current \eqref{eq:momentum_dens_diss} is given by
$P^{\mu}=(h,q^i)$ and, therefore, the boost velocity from an arbitrary
rest state ($u^i=0$) to the Landau rest frame ($u^i_L=0$) is given by
$v^i_L=q^i/h={\cal O}_1$. We follow the convention of
Refs.~\cite{1976AnPhy.100..310I,1979AnPhy.118..341I} to denote the
$n$-th order quantities in deviation from equilibrium by the symbol
${\cal O}_n$. The transformation of the charge currents into the
L-frame is then given by 
\bea\label{eq:N_i_trans}
N'^i_a = N^i_a-v^i_LN^0_a+{\cal O}_2.
\eea

Setting here $N^i_a=j^i_{a}$, $N'^i_a=j^i_{La}$ and
$N^0_a =n_a$, we find the charge diffusion fluxes in the
Landau rest state:
\bea\label{eq:diff_curr_Li}
{j}^{i}_{La}=
{j}^{i}_{a}-\frac{n_a}{h}q^i +{\cal O}_2.
\eea

It is worthwhile to note that the current on the left-hand side
${j}^{i}_{La}$ is evaluated at a transformed coordinate denoted by
$x'$. However, the difference
${j}^{i}_{La}(x')-{j}^{i}_{La}(x)\simeq (x'-x)\partial
{j}^{i}_{La}\propto v_L\partial {j}^{i}_{La}$ can be neglected, as it
is of the third order in gradients.   Because in the fluid rest frame
${j}^{0}_{La}=j^0_a=q^0=0$, we can express
Equation~\eqref{eq:diff_curr_Li} in a covariant form as
\bea
{j}^{\mu}_{La}
={j}^{\mu}_{a}-\frac{n_a}{h}q^\mu +{\cal O}_2,
\eea
which is valid already in an arbitrary frame, \ie, not exclusively in the fluid rest frame.
The~4-currents
\bea\label{eq:diff_curr1}
\mathscr{J}_a^\mu={j}^{\mu}_{a}-\frac{n_a}{h}q^\mu
\equiv {N}^{\mu}_{a}-\frac{n_a}{h}P^\mu
\eea
are the charge diffusion fluxes with respect to the energy flow, \ie,
these are the charge currents in the absence of energy flow. Note that
the combinations \eqref{eq:diff_curr1} are invariant under first-order
changes in $u^\mu$ despite the fact that the energy-diffusion flux
$q^\mu$ and the charge-diffusion fluxes $j^\mu_a$ depend on the chosen
velocity field.  It will be demonstrated below (see in
Section~\ref{sec:NS}), that the energy
dissipation in irreversible processes is associated precisely with
these currents.

The remaining thermodynamic variables which appear in
Equations~\eqref{eq:T_munu_decomp2} and \eqref{eq:N_a_decomp2} remain
unchanged if we neglect the second-order changes in thermodynamic
gradients induced by
$u^\mu$~\cite{1976AnPhy.100..310I,1979AnPhy.118..341I}.  Thus, we
can summarize the  relation between $u$ and $u_L$ as 
\bea\label{eq:u_Landau}
u^\mu_L =u^\mu +\frac{q^\mu}{h}+{\cal O}_2,
\eea
which follows straightforwardly upon comparing 
Equations~\eqref{eq:momentum_dens_diss} and \eqref{eq:vel_Landau1}.


The \emph{Eckart frame} (E-frame) is defined as the frame in which the velocity field
is  parallel to one of the conserved currents $N^\mu_a$.  For a fluid
with a single conserved charge (for example,
the net particle number), one has  $N^\mu=n u^\mu+j^\mu$ and, therefore,
the 4-velocity is defined~as~\cite{1940PhRv...58..919E}
\bea\label{eq:vel_Eckart}
u_E^\mu=\frac{N^\mu}{\sqrt{N^\mu N_\mu}}, 
\eea
which in combination with  Equations~\eqref{eq:match_cond}
and \eqref{eq:N_a_decomp2} gives
\bea\label{eq:n_Eckart}
n_E=\sqrt{N^\mu N_\mu},\qquad
N^\mu =n_E u_E^\mu,\qquad
j^\mu_E =0,
\eea
\ie, the particle diffusion flux vanishes.  The subscript $E$ in
Equations~\eqref{eq:vel_Eckart} and \eqref{eq:n_Eckart} indicates that the
quantities are evaluated in the E-frame.

The boost velocity from an arbitrary rest frame to the Eckart rest
frame is $v^i_E=j^i/n$. This implies that the velocities $u^\mu$ and
$u^\mu_E$ can be related by 
\bea\label{eq:u_Eckart}
u^\mu_E =u^\mu +\frac{j^\mu}{n}+{\cal O}_2.
\eea

We are now in a position to transform the energy flux into the Eckart
rest frame; we find 
\bea\label{eq:P_i_trans}
P'^{i}= P^{i}- \frac{j^i}{n}P^0+{\cal O}_2,
\eea

It follows then that the energy diffusion flux in the E-frame reads
\bea\label{eq:heat_curr_E}
{q}^{\mu}_{E}=
q^\mu-\frac{h}{n}{j}^{\mu}+{\cal O}_2.
\eea

The quantity
\bea\label{eq:heat_flux}
h^{\mu}=q^\mu-\frac{h}{n}{j}^{\mu}
=-\frac{h}{n}\mathscr{J}^\mu
\eea
can be seen as the energy flow with respect to the particle flow;
therefore, it can be identified with heat flux. The relation
\eqref{eq:heat_flux} demonstrates that {\it heat conduction and
  particle diffusion correspond to the same phenomenon, which is
  observed from different reference frames}. This is true only when
the higher than the first-order deviations from equilibrium can be
neglected.  To order ${\cal O}_1$, the relation
between the L-frame and the E-frame  can be established from
Equations~\eqref{eq:u_Landau} and \eqref{eq:u_Eckart} as
\bea\label{eq:u_EL}
u^\mu_L -u^\mu_E =
\frac{h^\mu}{h}=-\frac{\mathscr{J}^\mu}{n}.
\eea

Finally, let us note that the extension of the discussion above to
the case where multiple conserved charges are present is 
straightforward. This is achieved by attaching  a reference frame
to each of them, which formally then leads to the  definition
\bea\label{eq:vel_Eckart_a}
u_{a}^\mu=\frac{N^\mu_a}{\sqrt{N^\mu_a N_{a\mu}}}, 
\eea
which, in turn, enforces the vanishing of the corresponding diffusion flux, \ie,
$j^\mu_a=0$.  In what follows below,  we will keep the fluid
velocity generic (if not stated otherwise) without specifying any
particular reference frame.

\subsection{Equations of Relativistic Dissipative Hydrodynamics}

Equations of relativistic dissipative hydrodynamics are obtained by
substituting the decompositions \eqref{eq:T_munu_decomp2} and
\eqref{eq:N_a_decomp2} into the conservation laws
\eqref{eq:T_Na_cons}.  Using the same technique as in
Section~\ref{sec:ideal_hd} and recalling the
properties~\eqref{eq:orthogonality1}, we obtain
\bea \label{eq:hydro_1st_order1}
&& Dn_a +n_a\theta +\partial_\mu j^{\mu}_a=0,\\
\label{eq:hydro_1st_order2}
&& D\epsilon + (h+\Pi)\theta +\partial_\mu q^{\mu}-
q^{\mu}D u_\mu -\pi^{\mu\nu}\sigma_{\mu\nu}=0,\\
\label{eq:hydro_1st_order3}
&& (h+\Pi) D u_{\alpha}- \nabla_\alpha (p+\Pi)+ \Delta_{\alpha\mu}D q^{\mu}
+q^{\mu}\partial_\mu u_{\alpha}
+q_{\alpha}\theta +\Delta_{\alpha\nu}\partial_\mu \pi^{\mu\nu} =0,
\eea
where we introduced the velocity stress tensor as $\sigma_{\mu\nu}=\Delta_{\mu\nu}^{\alpha\beta}\partial_\alpha u_\beta$.

The system of
Equations~\eqref{eq:hydro_1st_order1}--\eqref{eq:hydro_1st_order3}
contains $l+4$ equations, as in the case of ideal hydrodynamics,
while the number of independent variables is now $4l+10$. The
additional unknown variables here are $3l$ components of the diffusion
fluxes, and 5 components of the shear stress tensor and the bulk viscous
pressure (recall that the equilibrium pressure is given by the
equation of state, and one of the diffusion fluxes
can always be eliminated). Thus, in order to solve the system
of Equations~\eqref{eq:hydro_1st_order1}--\eqref{eq:hydro_1st_order3}, we need
additional $3l+6$ equations for these dissipative quantities. These
relations are described in the phenomenological theory via the second
law of thermodynamics (the entropy principle).

\section{Relativistic Navier-Stokes (First-Order) Theory}
\label{sec:NS}

The entropy conservation law \eqref{eq:ent_gen_ideal} derived in the
framework of ideal fluid dynamics is no longer valid for dissipative
fluids. In this case, it should be replaced by the second law of
thermodynamics, which implies that the entropy production rate of an
isolated system must be always non-negative:
\bea\label{eq:second_law}
\partial_\mu S^\mu\ge 0,
\eea
where equality holds only for reversible processes.
By analogy with the decomposition of the charge currents
\eqref{eq:N_a_decomp2}, we can decompose $S^\mu$ into
 contributions parallel and orthogonal to $u^\mu$
\bea\label{eq:s_mu_decomp}
S^\mu_1 =su^\mu + j_s^\mu,
\eea
where $s$ is identified with the equilibrium entropy density (which is
sufficient for the first-order accuracy, as explained in
Section~\ref{sec:matching}), and the vector $j_s^\mu$ satisfies the
condition $u_\mu j_s^\mu=0$. The index 1 denotes
that~\eqref{eq:s_mu_decomp} is only the first-order approximation to
the entropy flux. For small departures from equilibrium, it is natural
to assume that $j_s^\mu$ is a linear combination of the energy and
charge diffusion fluxes
\bea\label{eq:entropy_diff}
j_s^\mu =\tilde{\beta} q^\mu -\sum\limits_a\tilde{\alpha}_a j_a^\mu,
\eea
where $\tilde{\beta}$ and $\tilde{\alpha}_a$ are functions of
thermodynamic variables and should be determined from the
condition~\eqref{eq:second_law}. This formulation of relativistic
dissipative fluid dynamics was proposed by
Eckart~\cite{1940PhRv...58..919E} and Landau--Lifshitz~\cite{Landau1987}, and leads to the relativistic version of
the NS theory. Inserting Equation~\eqref{eq:entropy_diff} into
Equation~\eqref{eq:s_mu_decomp} and introducing the dissipation function
via $R\equiv T\partial_\mu S^\mu $, we obtain
\bea\label{eq:ent_gen1}
R =
-\Pi\theta +\pi^{\mu\nu}\sigma_{\mu\nu}
+q^\mu (T\partial_\mu\tilde{\beta} +D u_\mu)
-T\sum\limits_a j_a^\mu\partial_\mu \tilde{\alpha}_a\nonumber\\
 +(T\tilde{\beta}-1) \partial_\mu q^{\mu}
+\sum\limits_a(\mu_a -T\tilde{\alpha}_a)\partial_\mu j_a^\mu,
\eea
where we employed the relations \eqref{eq:thermodyn1} and 
\eqref{eq:thermodyn3} and eliminated the terms $D\epsilon$ and
$Dn_a$ using Equations~\eqref{eq:hydro_1st_order1} and \eqref{eq:hydro_1st_order2}.
Requiring $R\geq 0$, we identify from Equation~\eqref{eq:ent_gen1} 
$\tilde{\beta}=\beta$, $\tilde{\alpha}_a = \alpha_a$. 
In the third and the fourth terms, we can replace $\partial_\mu\to\nabla_\mu$ 
due to the orthogonality conditions~\eqref{eq:orthogonality1}. Then, we have
\bea\label{eq:ent_gen11}
R =
-\Pi\theta +\pi^{\mu\nu}\sigma_{\mu\nu}+ q^\mu (T\nabla_\mu {\beta} +D u_\mu)
-T\sum\limits_aj_a^\mu\nabla_\mu {\alpha}_a.
\eea

We can further simplify the third term by approximating
$Du_\mu = h^{-1}\nabla_\mu p$ from Equation~\eqref{eq:ideal_hydro2} since
$q^\mu$ is already of the order ${\cal O}_1$.  Using
Equation~\eqref{eq:thermodyn6}, we obtain
\bea\label{eq:q_mod}
T\nabla_\mu\beta +Du_\mu =
T\sum\limits_a\frac{n_a}{h}\nabla_\mu \alpha_a.
\eea

Substituting this result into Equation~\eqref{eq:ent_gen11}
and recalling the definition \eqref{eq:diff_curr1}, we obtain finally
\bea\label{eq:ent_gen2}
R=-\Pi\theta +\pi^{\mu\nu}\sigma_{\mu\nu}
- T\sum\limits_a\mathscr{J}_a^\mu \nabla_\mu \alpha_a.
\eea

The second law of thermodynamics implies $R\geq 0$, which requires the
right-hand side of Equation~\eqref{eq:ent_gen2} to be a quadratic form of
thermodynamic forces $\theta$, $\pi^{\mu\nu}$ and
$\nabla_\mu \alpha_a$. Assuming linear dependence of the dissipative
fluxes on the thermodynamic forces, we obtain the constitutive
relations
\bea\label{eq:diss_currents}
\Pi =-\zeta \theta,\qquad
\pi^{\mu\nu}=2\eta \sigma^{\mu\nu},\qquad
\mathscr{J}_a^\mu =\sum\limits_b
\chi_{ab} \nabla^\mu \alpha_b,
\eea
where $\eta$ and $\zeta$ are called the shear and the bulk
viscosities, respectively, and $\chi_{ab}$ is the matrix of the
diffusion coefficients.  These relations together with
Equations~\eqref{eq:hydro_1st_order1}--\eqref{eq:hydro_1st_order3}
constitute a closed system of equations, which are known as
relativistic NS equations.

From Equations~\eqref{eq:ent_gen2} and \eqref{eq:diss_currents}, we obtain for the dissipation function
\bea\label{eq:ent_gen3}
R &=&
\zeta\theta^2 +2\eta \sigma^{\mu\nu}\sigma_{\mu\nu}
-T\sum\limits_{ab}\chi_{ab}\nabla^\mu \alpha_a\nabla_\mu \alpha_b\nonumber\\
&=& \frac{1}{\zeta}\Pi^2 +\frac{1}{2\eta} \pi^{\mu\nu}\pi_{\mu\nu}
-T\sum\limits_{ab}(\chi^{-1})_{ab}\mathscr{J}_a^\mu\mathscr{J}_{b\mu}.\quad
\eea

The positivity of this expression is guaranteed by the positivity of
the viscosity coefficients and the eigenvalues of the matrix
$\chi_{ab}$ (recall that the diffusion fluxes $\mathscr{J}_a^\mu$ are
spatial). We see from Equation~\eqref{eq:ent_gen3} that the contribution of
diffusion processes to the function $R$ depends only on the relative
diffusion currents $\mathscr{J}_{a}^{\mu}$, but not on the currents
$j_a^\mu$ and $q^\mu$ separately. This result, which was  obtained in the
framework of the relativistic NS (first-order) theory, is the direct
consequence of the Lorentz invariance and indicates that the
dissipation in the fluid is independent of the choice of the fluid
velocity field, as expected.

The entropy current given by Equations~\eqref{eq:s_mu_decomp} and \eqref{eq:entropy_diff}
can now be written as
\bea\label{eq:s_mu}
S^\mu_1 =su^\mu + \beta q^\mu - 
\sum\limits_a\alpha_aj_a^\mu 
=\frac{s}{h}P^\mu  - \sum\limits_a
\alpha_a\mathscr{J}_a^\mu,
\eea
where we used Equations~\eqref{eq:thermodyn3}, \eqref{eq:beta_alpha},
\eqref{eq:momentum_dens_diss} and \eqref{eq:diff_curr1} to obtain the
second relation. One can give a simple interpretation of the second
expression in Equation~\eqref{eq:s_mu}. Recalling that $P^\mu/h= u^\mu_L$
is the fluid velocity measured in the L-frame (see
Equation~\eqref{eq:momentum_dens_diss}), we observe that the first term on
the right-hand side of Equation~\eqref{eq:s_mu} is the entropy current that
is convected together with the energy.  The second term arises as a
result of the relative flow between the energy and the charges and,
therefore, should be identified with the irreversible part of the
entropy flow.  Using Equations~\eqref{eq:thermodyn3},
\eqref{eq:beta_alpha}, \eqref{eq:momentum_dens_diss} and
\eqref{eq:diff_curr1}, we can rewrite Equation~\eqref{eq:s_mu} also in the following form:
\bea\label{eq:s_mu_first_order}
S^\mu_1 =\beta P^\mu-
\sum\limits_a\alpha_a N^\mu_{a}
=p\beta^\mu+\beta_\nu T^{\mu\nu}-
\sum\limits_a\alpha_a N^\mu_{a},
\eea
which formally coincides with its counterpart of ideal hydrodynamics \eqref{eq:s_mu_ideal1}.

If we have only one sort of conserved charge ($l=1$), then
instead of the third relation in Equation~\eqref{eq:diss_currents}, we have simply
\bea\label{eq:heat_currents1}
\mathscr{J}^\mu =
\chi\nabla^\mu \alpha,\qquad
h^\mu =-
\kappa \frac{nT^2}{h}\nabla^\mu\alpha, 
\eea
where we used the definition of the heat current given by Equation~\eqref{eq:heat_flux} and
introduced the coefficient of thermal conductivity via
\bea\label{eq:kappa_chi}
\kappa=\left(\frac{h}{nT}\right)^2\chi.
\eea

Equation \eqref{eq:kappa_chi} establishes the relation between 
the thermal conductivity and the diffusion coefficient $\chi$. 
Thus, there are only three independent transport coefficients 
in the first-order theory which relate the irreversible fluxes 
to the corresponding thermodynamic forces. Employing Equation~\eqref{eq:q_mod}, 
we can write the heat flux~\eqref{eq:heat_currents1} in the following way
\bea\label{eq:Eckart_current}
h^\mu 
=\kappa \left(\nabla^\mu T -\frac{T}{h}\nabla^\mu p\right),
\eea
which in the fluid rest frame reads
\bea\label{eq:Eckart_current_rest}
h^i =-\kappa \left(\partial_i T -\frac{T}{h}\partial_i p\right).
\eea

Equation \eqref{eq:Eckart_current_rest} is the relativistic
generalization of the well-known Fourier law
$h^i=-\kappa \partial_i T$ of the non-relativistic
hydrodynamics~\cite{Landau1987}.

The expressions \eqref{eq:ent_gen3}--\eqref{eq:s_mu_first_order} in the case of $l=1$
reduce to
\bea\label{eq:ent_gen4}
R =
\zeta\theta^2 +2\eta \sigma^{\mu\nu}\sigma_{\mu\nu}
-T\chi\left(\nabla^\mu \alpha\right)^2 =
\frac{1}{\zeta}\Pi^2 +\frac{1}{2\eta} \pi^{\mu\nu}\pi_{\mu\nu}
-\frac{1}{\kappa T}h^\mu h_\mu,
\eea
and 
\bea\label{eq:s_mu1}
S^\mu_1 =su^\mu + \beta q^\mu-\alpha j^\mu 
 =\frac{s}{n}N^\mu + \beta h^\mu.
\eea

The last relation in Equation~\eqref{eq:s_mu1} is the 
decomposition of the entropy flow into its reversible and 
irreversible components observed from the E-frame.

Note that in the case where there are no conserved charges, 
\ie, when $l=0$, the heat conduction and/or diffusion phenomena 
are absent~\cite{1985PhRvD..31...53D}.

\section{Israel--Stewart (Second-Order) Theory}
\label{sec:IS}

The first-order theory described in the previous section turns out to
be acausal, and, therefore, cannot be regarded as a consistent theory
of relativistic dissipative fluids. The origin of acausality lies in
the constitutive relations \eqref{eq:diss_currents}, which imply that
the thermodynamic forces generate dissipative fluxes
instantaneously~\cite{1967ZPhy..198..329M,1976AnPhy.100..310I,
  1983AnPhy.151..466H,1987PhRvD..35.3723H}. In addition, this theory
suffers also from instability, which is a consequence of
acausality~\cite{1983AnPhy.151..466H,1985PhRvD..31..725H,
  1987PhRvD..35.3723H,Denicol2008JPhG,Pu2010PhRvD}. However, as
pointed out in the introduction, acausalities, and instabilities are a
consequence of the matching procedure to the local-equilibrium
reference state which can be generalized to obtain causal and stable
first-order dissipative
hydrodynamics~\cite{Bemfica:2019knx,Kovtun:2019hdm,Noronha2022PhRvL}.

It turns out that to recover the causality, the entropy 
current $S^\mu$ is required to be at least a quadratic function 
of the dissipative fluxes. This idea of an extension of the entropy 
current up to the second order was first proposed by M\"uller~\cite{1967ZPhy..198..329M} 
for nonrelativistic fluids. For relativistic fluids, a similar second-order 
theory was developed by Israel and Stewart~\cite{1976AnPhy.100..310I,1979AnPhy.118..341I}.
In this subsection, we review briefly the Israel--Stewart (IS) formulation 
of causal hydrodynamics, following mainly Ref.~\cite{1976AnPhy.100..310I}.

As a starting point, the entropy current given by
Equations~\eqref{eq:s_mu} and \eqref{eq:s_mu_first_order} is extended
up to the second order in dissipative quantities $j^\mu_a$, $q^\mu$,
$\pi^{\mu\nu}$ and $\Pi$.  It is worth stressing that, despite the
frame dependence of these quantities, the entropy current $S^\mu$
along with the energy--momentum tensor $T^{\mu\nu}$ and the charge
currents $N_a^\mu$ should be regarded as a {\it primary variable} and
should be therefore frame-independent up to the second
order~\cite{1976AnPhy.100..310I,1979AnPhy.118..341I} (we note that the
expressions \eqref{eq:s_mu} and \eqref{eq:s_mu_first_order} are
frame-independent only at the first order in deviations from
equilibrium).

We write now the entropy current in the following form:
\bea\label{eq:s_mu_IS}
S^\mu ={S}^\mu_1 -\bar{R}^\mu- R^\mu, 
\eea
where ${S}^\mu_1$ is the first-order contribution given by 
Equations~\eqref{eq:s_mu} and \eqref{eq:s_mu_first_order}, and the 
terms $\bar{R}^\mu$ and $R^\mu$ collect all possible second-order corrections. 

The most general form of the vectors $R^\mu$ and $\bar{R}^\mu$ in a generic frame is~\cite{1976AnPhy.100..310I}
\bea\label{eq:vector_R}
R^\mu &=& \frac{\beta u^\mu}{2}\bigg(\beta_\Pi \Pi^2+\beta_\pi \pi^{\alpha\beta}\pi_{\alpha\beta}-\sum\limits_{ab}\beta_{\mathscr{J}}^{ab}
\mathscr{J}^\alpha_{a}\mathscr{J}_{b\alpha}\bigg)\nonumber\\
&&+ \beta\Pi\sum\limits_a\alpha_\Pi^a \mathscr{J}^\mu_{a}+
\beta\pi^\mu_\nu\sum\limits_a\alpha_\pi^a \mathscr{J}^\nu_{a},\quad\\
\label{eq:vector_R_bar}
\bar{R}^\mu &=& \beta h^{-1}\bigg(\Pi q^\mu-\pi^{\mu\nu}q_\nu-
\frac{1}{2}u^\mu q^\nu q_\nu\bigg),
\eea 
where the new coefficients $\beta_\Pi$, $\beta_\pi$, 
$\beta_{\mathscr{J}}^{ab}= \beta_{\mathscr{J}}^{ba}$, $\alpha_\Pi^a$ 
and $\alpha_\pi^a$ are unknown functions of $\epsilon$ and $n_a$. The 
vector $R^\mu$ is frame-independent up to the second order, and 
$\bar{R}^\mu$ collects the second-order contributions to $S^\mu$ 
which are not frame-independent to the order ${\cal O}_2$
(note that we use different metric signature from Ref.~\cite{1976AnPhy.100..310I}).
Because the contribution ${S}^\mu_1$ is frame-independent only 
to the first order, the term $\bar{R}^\mu$ is necessary to provide 
the frame-independence of the total entropy current \eqref{eq:s_mu_IS}~\cite{1976AnPhy.100..310I}.

 The terms in Equations~\eqref{eq:vector_R} and \eqref{eq:vector_R_bar} 
 which are proportional to $u^\mu$ are responsible for the second-order 
 corrections to the equilibrium entropy density $s$. Indeed, the 
 non-equilibrium entropy density is identified with $S=S^\mu u_\mu$, 
 which can be found from \linebreak Equations~\eqref{eq:s_mu_IS}--\eqref{eq:vector_R_bar}
\bea\label{eq:ent_dens_IS}
S=
s-\frac{\beta}{2}\left(\beta_\Pi \Pi^2+\beta_\pi 
\pi^{\alpha\beta}\pi_{\alpha\beta}-\sum\limits_{ab}
\beta_{\mathscr{J}}^{ab}\mathscr{J}^\alpha_{a}
\mathscr{J}_{b\alpha}- h^{-1}q^\alpha q_\alpha\right)\leq s,
\eea
where we used Equations~\eqref{eq:s_mu}. The last inequality in 
Equation~\eqref{eq:ent_dens_IS} requires the entropy density in 
non-equilibrium states to be smaller than its equilibrium 
value $s$. From here, we conclude that $\beta_\Pi\geq 0$, 
$\beta_\pi\geq 0$, and  the matrix $\beta_{\mathscr{J}}^{ab}$ 
is positive-semidefinite. The term in Equation~\eqref{eq:ent_dens_IS},
which is proportional to $q^\mu q_\mu$ represents the shift in 
the entropy density because of the change of the reference frame 
and is automatically negative. Those terms in Equations~\eqref{eq:vector_R} 
and \eqref{eq:vector_R_bar} which are orthogonal to $u^\mu$ 
represent the irreversible entropy flux arising from couplings 
between the diffusion and viscous fluxes.
 
 The phenomenological equations for the dissipative fluxes should 
 be found again from the positivity condition of the dissipative function
\bea\label{eq:second_law_IS}
R=T\partial_\mu S^\mu =T\partial_\mu {S}^\mu_1- 
T\partial_\mu \bar{R}^\mu -T\partial_\mu R^\mu\geq 0.
\eea

 The first term in Equation~\eqref{eq:second_law_IS} was 
 already computed in Equation~\eqref{eq:ent_gen11}:
\bea\label{eq:S_bar_deriv}
T\partial_\mu {S}^\mu_1 =
-\Pi\theta +\pi^{\mu\nu}\sigma_{\mu\nu}+ q^\mu (T\nabla_\mu {\beta} +D u_\mu)
-T\sum\limits_aj_a^\mu\nabla_\mu {\alpha}_a.
\eea

 Now we use Equation~\eqref{eq:hydro_1st_order3} to eliminate 
 the acceleration term in Equation~\eqref{eq:S_bar_deriv}: 
\bea\label{eq:Du_modify_IS}
h D u_{\mu} = \nabla_\mu p+ \nabla_\mu\Pi
-\Pi D u_\mu - \Delta_{\mu\nu}D q^{\nu}
-q^{\nu}\partial_\nu u_{\mu}
-q_{\mu}\theta -\Delta_{\mu\nu}\partial_\alpha \pi^{\alpha\nu}=-hT\nabla_\mu \beta \nonumber\\
+ 
T\sum\limits_a n_a\nabla_\mu \alpha_a+
\nabla_\mu\Pi
-\Pi D u_\mu - \Delta_{\mu\nu}D q^{\nu}
-q^{\nu}\partial_\nu u_{\mu}
-q_{\mu}\theta -\Delta_{\mu\nu}\partial_\alpha \pi^{\alpha\nu},
\eea
where we used the second relation of Equation~\eqref{eq:thermodyn6} 
to modify the pressure gradient in Equation~\eqref{eq:Du_modify_IS}.  
Equation \eqref{eq:Du_modify_IS} differs from Equation~\eqref{eq:q_mod} 
by additional second-order terms, which now cannot be neglected. 
Substituting Equation~\eqref{eq:Du_modify_IS} in the third term of 
Equation~\eqref{eq:S_bar_deriv} and recalling the definitions~\eqref{eq:diff_curr1}, we obtain
\bea\label{eq:S_bar_deriv1}
T\partial_\mu {S}^\mu_1
&=&-\Pi\theta +\pi^{\mu\nu}\sigma_{\mu\nu}
-T\sum\limits_a\mathscr{J}_a^\mu\nabla_\mu {\alpha}_a
+h^{-1} \big(q^{\mu}\partial_\mu\Pi
\nonumber\\
&&-\Pi q^{\mu}D u_\mu  - q_{\mu}D q^{\mu}
-q^{\mu}q^{\nu}\partial_\nu u_{\mu}
-q^{\mu}q_{\mu}\theta -q_{\nu}\partial_\mu \pi^{\mu\nu}\big),
\eea
where we used the notation $\nabla_\mu=\Delta_{\mu\nu}\partial^\nu$.
Using Equations~\eqref{eq:vector_R_bar} and \eqref{eq:S_bar_deriv1}, we obtain
\bea\label{eq:SR_bar_deriv}
T\partial_\mu {S}^\mu_1 -T\partial_\mu \bar{R}^\mu 
&=&-\Pi\Big[\theta +h^{-1} \partial_\mu q^{\mu} + 
\bar{\gamma}_\Pi h^{-1}q^{\mu} Du_\mu + \tilde{\gamma}_\Pi 
Tq^{\mu}\partial_\mu (\beta h^{-1}) \Big]\nonumber\\
&+& \pi^{\mu\nu}\Big[\sigma_{\mu\nu} +h^{-1} \partial_\mu q_\nu
 +\bar{\gamma}_\pi Tq_\nu \partial_\mu (\beta h^{-1})\Big]-
 T\sum\limits_a\mathscr{J}_a^\mu\nabla_\mu {\alpha}_a\nonumber\\
&-&
 q^{\mu}\bigg[h^{-1}q^{\nu}\partial_\nu u_{\mu} +\frac{1}{2}h^{-1}q_{\mu}\theta 
-\frac{1}{2}T q_\mu  D(\beta h^{-1})
+(1-\bar{\gamma}_\Pi)h^{-1} \Pi  Du_\mu \nonumber\\
&+&  
(1-\tilde{\gamma}_\Pi)T\Pi \partial_\mu (\beta h^{-1})
- (1-\bar{\gamma}_\pi) T\pi_\mu^\nu \partial_\nu (\beta h^{-1})\bigg].
\eea

Here, we introduced three additional coefficients $\bar{\gamma}_\Pi$, 
$\tilde{\gamma}_\Pi$ and $\bar{\gamma}_\pi$ because there is an 
ambiguity in ``sharing'' the terms involving $\Pi q^{\mu}$ and 
$\pi^\mu_\nu q^\nu$ between the diffusion and viscous fluxes~\cite{1983AnPhy.151..466H}.
Note that Ref.~\cite{1976AnPhy.100..310I} assumed $\bar{\gamma}_\pi=\bar{\gamma}_\Pi=
\tilde{\gamma}_\Pi=0$. 

Now we take the divergence of Equation~\eqref{eq:vector_R}:
\bea\label{eq:R_deriv}
\partial_\mu {R}^\mu 
&=& \beta\Pi\bigg[\beta_\Pi D\Pi
+\frac{1}{2} T\Pi\partial_\mu (\beta \beta_\Pi u^\mu)  +
\sum\limits_a\alpha_\Pi^a \partial_\mu \mathscr{J}^\mu_{a}+
\gamma_\Pi T\sum\limits_a\mathscr{J}^\mu_{a} \partial_\mu(\beta\alpha_\Pi^a)\bigg]\nonumber\\
&+& \beta \pi^{\mu\nu}\bigg[\beta_\pi D \pi_{\mu\nu}+
\frac{1}{2}T \pi_{\mu\nu}\partial_\lambda (\beta \beta_\pi u^\lambda)
+\sum\limits_a\alpha_\pi^a \partial_\mu \mathscr{J}_{a\nu}+\gamma_\pi T\sum\limits_a \mathscr{J}_{a\nu}\partial_\mu (\beta \alpha_\pi^a)\bigg]\nonumber\\
&-& \beta \sum\limits_a \mathscr{J}^\mu_{a}\bigg[\sum\limits_{b}
\beta_{\mathscr{J}}^{ab}D\!\!\mathscr{J}_{b\mu}
+\frac{1}{2}T\sum\limits_{b}\mathscr{J}_{b\mu}\partial_\nu
(\beta \beta_{\mathscr{J}}^{ab}u^\nu) 
-\alpha_\Pi^a \partial_\mu \Pi
-\alpha_\pi^a \partial_\nu \pi^\nu_\mu\nonumber\\
&-& (1-\gamma_\Pi)T\Pi\partial_\mu(\beta \alpha_\Pi^a)
-(1-\gamma_\pi)T\pi^\nu_\mu \partial_\nu (\beta \alpha_\pi^a)\bigg],\quad
\eea
where we introduced again two additional coefficients $\gamma_\Pi$ and 
$\gamma_\pi$ to ``share'' the terms containing $\Pi \mathscr{J}_{a}^{\mu}$ 
and $\pi^\mu_\nu \mathscr{J}_{a}^{\nu}$ between the diffusion and viscous 
fluxes~\cite{1983AnPhy.151..466H}. We kept also all terms that contain 
gradients of transport coefficients which were neglected in 
Ref.~\cite{1976AnPhy.100..310I}.
 
Combining Equations~\eqref{eq:SR_bar_deriv} and \eqref{eq:R_deriv}, 
we obtain for the dissipative function \eqref{eq:second_law_IS}
\bea\label{eq:R_IS_Landau}
R &=&-\Pi\bigg[\theta +\beta_\Pi \dot{\Pi} +
\sum\limits_a \alpha_\Pi^a \partial_\mu \mathscr{J}^\mu_{a}
+\gamma_\Pi T\sum\limits_a\mathscr{J}^\mu_{a} \partial_\mu(\beta\alpha_\Pi^a)
+\frac{1}{2} T \Pi\partial_\mu (\beta\beta_\Pi u^\mu)\nonumber\\
&& \hspace{4cm} +h^{-1} \partial_\mu q^{\mu} + 
\bar{\gamma}_\Pi h^{-1}q^{\mu} \dot{u}_\mu + \tilde{\gamma}_\Pi 
Tq^{\mu}\partial_\mu (\beta h^{-1})\bigg]\nonumber\\
&+&\pi^{\mu\nu}\bigg[ \sigma_{\mu\nu} -\beta_\pi \dot{\pi}_{\mu\nu}
-\sum\limits_a \alpha_\pi^a \nabla_{<\mu} \mathscr{J}_{a\nu >}- 
\gamma_\pi T\sum\limits_a \mathscr{J}_{a<\nu}\nabla_{\mu>}
(\beta \alpha_\pi^a) \nonumber\\
&& \hspace{2.2cm}  -
\frac{1}{2} T\pi_{\mu\nu}\partial_\lambda (\beta\beta_\pi u^\lambda) 
 +h^{-1} \partial_{< \mu} q_{\nu >} +\bar{\gamma}_\pi Tq_{<\nu} 
\partial_{\mu >}(\beta h^{-1})\bigg]\nonumber\\ 
&-&\sum\limits_a \mathscr{J}^\mu_{a}\bigg[T\nabla_\mu {\alpha}_a-\sum\limits_{b}\beta_{\mathscr{J}}^{ab}\dot{\mathscr{J}}_{b\mu}
-\frac{1}{2}T\sum\limits_{b}\mathscr{J}_{b\mu}\partial_\nu
(\beta\beta_{\mathscr{J}}^{ab}u^\nu)+\alpha_\Pi^a \nabla_\mu \Pi \nonumber\\
&&\hspace{0.5cm}+\alpha_\pi^a \Delta_{\alpha\mu}\partial_\nu \pi^{\nu\alpha} +(1-\gamma_\Pi)T\Pi\nabla_\mu(\beta \alpha_\Pi^a)
+(1-\gamma_\pi)T\pi^\nu_\mu\partial_\nu(\beta \alpha_\pi^a)\bigg]
\qquad\qquad\nonumber\\
 &-&
 q^{\mu}\bigg[h^{-1}q^{\nu}\partial_\nu u_{\mu} 
+\frac{1}{2}h^{-1}q_{\mu}\theta 
-\frac{1}{2}T q_\mu  D(\beta h^{-1})
+(1-\bar{\gamma}_\Pi)h^{-1} \Pi \dot{u}_\mu \nonumber\\
&&\hspace{.5cm}+  
(1-\tilde{\gamma}_\Pi)T\Pi \nabla_\mu (\beta h^{-1})
- (1-\bar{\gamma}_\pi) T\pi_\mu^\nu 
\nabla_\nu (\beta h^{-1})\bigg],
\eea
where we introduced the short-hand notations
\bea\label{eq:dot_def}
\dot{u}_\mu = Du_\mu,\qquad
\dot{\Pi}=D\Pi,\qquad
\dot{\pi}_{\mu\nu}=\Delta_{\mu\nu\rho\sigma}
D{\pi}^{\rho\sigma},\qquad
\mathscr{\dot{J}}_{a\mu}=\Delta_{\mu\nu} D\!\!\mathscr{J}^\nu_a,\\
\label{eq:A_mu_nu}
A_{<\mu\nu>}=\Delta_{\mu\nu}^{\alpha\beta}A_{\alpha\beta},\hspace{4.5cm}
\eea
and used the symmetries of the dissipative fluxes.

The expressions in the square brackets in Equation~\eqref{eq:R_IS_Landau} 
are called {\it generalized or extended thermodynamic forces}.
Requiring $R\geq 0$, fixing for simplicity the L-frame $q^\mu=0$
 and assuming linear relations between these 
forces and the dissipative fluxes, we obtain the following evolution equations:
\bea\label{eq:bulk_IS}
\Pi &=& -\zeta\bigg[\theta +\beta_\Pi \dot{\Pi}  +
\sum\limits_a \alpha_\Pi^a \partial_\mu \mathscr{J}^\mu_{a}
+\gamma_\Pi T\sum\limits_a\mathscr{J}^\mu_{a} \partial_\mu(\beta\alpha_\Pi^a)\nonumber\\
&&\hspace{0.5cm}+\frac{1}{2} T \Pi \partial_\mu(\beta\beta_\Pi u^\mu)\bigg],\\
\label{eq:shear_IS}
\pi_{\mu\nu} &=& 2\eta\bigg[ \sigma_{\mu\nu} -\beta_\pi \dot{\pi}_{\mu\nu}-\sum\limits_a \alpha_\pi^a \nabla_{<\mu} \mathscr{J}_{a\nu >}
  \nonumber\\
  &&\hspace{0.5cm} -\gamma_\pi T\sum\limits_a \mathscr{J}_{a<\nu}\nabla_{\mu >}(\beta \alpha_\pi^a) 
-\frac{1}{2} T\pi_{\mu\nu} \partial_\lambda(\beta\beta_\pi u^\lambda) \bigg],\qquad\\
\label{eq:current_IS}
\mathscr{J}_{a\mu} & =& \sum\limits_b \chi_{ab}\bigg[\nabla_\mu {\alpha}_b-\beta \sum\limits_{c}\beta_{\mathscr{J}}^{bc}\dot{\mathscr{J}}_{c\mu}
-\frac{1}{2}\sum\limits_{c}\mathscr{J}_{c\mu}\partial_\nu(\beta\beta_{\mathscr{J}}^{bc}u^\nu)  
+ \beta\alpha_\Pi^b \nabla_\mu \Pi\nonumber\\
&& \hspace{0.5cm}  
 +\beta\alpha_\pi^b \Delta_{\mu\alpha}\partial_\nu \pi^{\nu\alpha} +(1-\gamma_\Pi)\Pi\nabla_\mu(\beta \alpha_\Pi^b)
+(1-\gamma_\pi)\pi^\nu_\mu\partial_\nu(\beta \alpha_\pi^b)\bigg].
\eea
\textls[-15]{Equations \eqref{eq:bulk_IS}--\eqref{eq:current_IS} in this form for
{\it single type of conserved charges} were derived in
Ref.~\cite{1983AnPhy.151..466H} and were written in the E-frame. In
the original papers of Israel and Stewart
\cite{1976AnPhy.100..310I,1976PhLA...58..213I,1979AnPhy.118..341I}, the
terms which are nonlinear in thermodynamic forces and dissipative
fluxes were dropped, {although the general case of chemically
  reacting multicomponent mixtures was discussed in
  Ref.~\cite{1976AnPhy.100..310I}. The current form generalizes the
  hydrodynamics equations derived in
  Refs}.~\cite{1976AnPhy.100..310I,1976PhLA...58..213I,1979AnPhy.118..341I,1983AnPhy.151..466H}
  to the case of multiple independent flavors of conserved charges,
  where all second-order terms arising from the entropy current are
  kept.}

With the aid of Equations~\eqref{eq:bulk_IS}--\eqref{eq:current_IS}, the dissipative function \eqref{eq:R_IS_Landau} obtains the form
\bea\label{eq:R_IS_final}
R = \frac{1}{\zeta}\Pi^2 +\frac{1}{2\eta} 
\pi^{\mu\nu}\pi_{\mu\nu}
-T\sum\limits_{ab} \mathscr{J}_a^\mu (\chi^{-1})_{ab}\mathscr{J}_{b\mu},
\eea
which formally coincides with Equation~\eqref{eq:ent_gen3}.
Defining relaxation times according to 
\bea\label{eq:IS_relax_times}
\tau_\Pi =\zeta \beta_\Pi,\qquad
\tau_\pi =2\eta\beta_\pi,\qquad
\tau_\mathscr{J}^{ac} =\beta \sum\limits_b\chi_{ab}\beta_{\mathscr{J}}^{bc},
\eea
we can write Equations~\eqref{eq:bulk_IS}--\eqref{eq:current_IS} in the following form:
\bea\label{eq:bulk_IS_relax}
\tau_\Pi\dot{\Pi}+\Pi &=&-\zeta\theta -\zeta\bigg[\sum\limits_a 
\alpha_\Pi^a \partial_\mu \mathscr{J}^\mu_{a}
+\gamma_\Pi T\sum\limits_a\mathscr{J}^\mu_{a} \partial_\mu(\beta\alpha_\Pi^a)
\nonumber\\
&&\hspace{1cm}
+\frac{1}{2} T \Pi \partial_\mu(\beta\beta_\Pi u^\mu)\bigg],\\
\label{eq:shear_IS_relax}
\tau_\pi\dot{\pi}_{\mu\nu}+\pi_{\mu\nu}&=& 2\eta \sigma_{\mu\nu}
-2\eta\bigg[ \sum\limits_a \alpha_\pi^a \nabla_{<\mu} \mathscr{J}_{a\nu >}+ 
\gamma_\pi T\sum\limits_a \mathscr{J}_{a<\nu}\nabla_{\mu >}(\beta \alpha_\pi^a) 
\nonumber\\
&&\hspace{1cm}
+\frac{1}{2} T\pi_{\mu\nu} \partial_\lambda(\beta\beta_\pi u^\lambda)\bigg],\\
\label{eq:current_IS_relax}
\sum_{c}\tau^{ac}_\mathscr{J}
\dot{\mathscr{J}}_{c\mu}+\mathscr{J}_{a\mu} &=& \sum\limits_b \chi_{ab}\nabla_\mu {\alpha}_b+\sum\limits_b \chi_{ab}\bigg[
-\frac{1}{2}\sum\limits_{c}\mathscr{J}_{c\mu}\partial_\nu(\beta\beta_{\mathscr{J}}^{bc}u^\nu)  
+ \beta\alpha_\Pi^b \nabla_\mu \Pi \nonumber\\
&+&\beta\alpha_\pi^b \Delta_{\mu\alpha}\partial_\nu \pi^{\nu\alpha} 
+(1-\gamma_\Pi)\Pi\nabla_\mu(\beta \alpha_\Pi^b)
+(1-\gamma_\pi)\pi^\nu_\mu\partial_\nu(\beta \alpha_\pi^b)\bigg].\quad
\eea
The first terms on the right-hand sides of these equations 
represent the corresponding NS contributions to the dissipative 
fluxes; see Equation~\eqref{eq:diss_currents}. The first terms on the 
left-hand sides incorporate the relaxation of the dissipative 
fluxes to their NS values on finite time scales given by 
Equation~\eqref{eq:IS_relax_times}. Thus, these relaxation terms imply 
a delay in the response of the dissipative fluxes to thermodynamic 
forces and recover the causality of the theory~\cite{Denicol2008JPhG,Pu2010PhRvD}.
The rest of the terms in Equations~\eqref{eq:bulk_IS_relax}--\eqref{eq:current_IS_relax} 
are responsible for spatial inhomogeneities in the dissipative 
fluxes as well as nonlinear couplings between different dissipative processes. 

We note that the derivation of the second-order hydrodynamics from 
the kinetic theory produces additional terms which are not obtained 
within the phenomenological theory~\cite{1979AnPhy.118..341I}. The 
derivation of complete IS equations from kinetic theory is discussed in Refs.~\cite{2009PrPNP..62..556B,Betz2011EPJWC,2012EPJA...48..170D,
Denicol2012PhRvD,2014JPhG...41l4004D}.

In the case of one conserved current, we have instead of Equations~\eqref{eq:bulk_IS_relax}--\eqref{eq:current_IS_relax}
\bea\label{eq:bulk_IS_relax1}
\tau_\Pi\dot{\Pi}+\Pi &=&-\zeta\theta -\zeta\bigg[\alpha_\Pi 
\partial_\mu \mathscr{J}^\mu +\gamma_\Pi T\!\mathscr{J}^\mu 
\partial_\mu(\beta\alpha_\Pi)+\frac{1}{2} T \Pi \partial_\mu
(\beta\beta_\Pi u^\mu)\bigg],\\ 
\label{eq:shear_IS_relax1}
\tau_\pi\dot{\pi}_{\mu\nu}+\pi_{\mu\nu}&=& 2\eta \sigma_{\mu\nu}-
2\eta\bigg[ \alpha_\pi \nabla_{<\mu} \mathscr{J}_{\nu >}+ \gamma_\pi 
T\! \mathscr{J}_{<\nu}\nabla_{\mu >}(\beta \alpha_\pi)\nonumber\\
&&\hspace{0.5cm}+\frac{1}{2} 
T\pi_{\mu\nu} \partial_\lambda(\beta\beta_\pi u^\lambda)\bigg],\\
\label{eq:current_IS_relax1}
\tau_{\!\mathscr{J}}\!\dot{\mathscr{J}}_{\mu}+\mathscr{J}_{\mu} &=& 
\chi\nabla_\mu {\alpha}+\chi\bigg[ \beta\alpha_\Pi \nabla_\mu \Pi
+\beta\alpha_\pi \Delta_{\mu\alpha}\partial_\nu \pi^{\nu\alpha} -\frac{1}{2}\mathscr{J}_{\mu}\partial_\nu(\beta\beta_{\!\mathscr{J}}u^\nu)  
\nonumber\\
&& \hspace{0.5cm} +(1-\gamma_\Pi)\Pi\nabla_\mu(\beta \alpha_\Pi)
+(1-\gamma_\pi)\pi^\nu_\mu\partial_\nu(\beta \alpha_\pi)\bigg],
\eea
where $\tau_{\!\mathscr{J}} =\beta \chi\beta_{\!\mathscr{J}}$.

\section{Summary}
\label{sec:summary}

We provided a review of the phenomenological theory of second-order
relativistic hydrodynamics for systems with multiple conserved charges
with an extension to multiflavor fluids. The hydrodynamic state of the
system is described through the energy--momentum tensor and the
4-currents of conserved charges. We reviewed the derivation and
content of the equations at zeroth, first, and second order in
gradient expansion of the energy--momentum tensor and currents, which
led us to the ideal Navier--Stokes and Israel--Stewart hydrodynamics,
respectively.  From the positivity condition of the dissipative
function, the most general set of dissipative processes was identified
which contains also the relative diffusions between different
conserved currents. { We kept all the second-order gradient terms
  arising from the second law of thermodynamics, as well as the
  nonlinear terms in the thermodynamic gradients and dissipative
  fluxes, which were omitted in the canonical Israel--Stewart theory,
  but were kept in Ref.~\cite{1983AnPhy.151..466H} in the case of one
  conserved flavor.}

The hydrodynamics theory exposed in this article is phenomenological
in nature. More formal but also complex derivations are available in
the literature which utilize different concepts and approaches of
statistical mechanics, for example, quasiparticle Boltzmann
equation~\cite{Denicol2012PhRvD,Denicol2012PhLB,Molnar2016PhRvD} or
non-equilibrium statistical
operator~\cite{Harutyunyan:2018cmm,Harutyunyan:2021rmb}. Nevertheless,
the phenomenological theory can be reasonably expected to remain in
the arsenal of theoretical tools that can be deployed in studies of
fluid in new settings and/or under various external fields.

{ A few closing remarks are in order concerning
  the numerical implementations of the second-order dissipative
  hydrodynamics. First of all, the choice between the frames (\ie,
  Eckart vs. Landau) may be significantly influenced by the required
  computational cost, which in turn depends on such factors as system
  size, boundary conditions and the level of accuracy required. In
  general, the Landau frame is known to be computationally more
  expensive compared to the Eckart frame, but it also provides more
  insight into the behavior of the fluid. Eventually, the choice
  between the Landau and Eckart frames will depend on the specific
  problem at hand and the trade-off between accuracy and computational
  efficiency.  A number of methods are available for the solution of
  the generalized (second-order) Navier--Stokes equations for
  multifluids, for example, finite difference method, finite volume
  method, spectral methods, and Lagrangian particle methods; see
  Ref.~\cite{Heinz2015IJMPE} Section 5 for a pedagogical discussion
  in the context of relativistic heavy ion collisions.  Again, the
  choice of the method will be dictated by the specifics of the problem
  at hand, the computational cost, and required accuracy. }
\vspace{12pt}

\authorcontributions{A. H. and A. S. equally contributed to this research. 
}

\funding{A. H. and A. S. were funded by Volkswagen Foundation (Hannover,
Germany) grant No. 96 839.  A.~S. was funded by
Deutsche Forschungsgemeinschaft Grant No. SE 1836/5-2 and the Polish
NCN Grant No. 2020/37/B/ST9/01937 at Wroc\l{}aw University. 
}



\conflictsofinterest{The authors declare no conflict of interest 
} 

\begin{adjustwidth}{-\extralength}{0cm}

\reftitle{References 
}

\PublishersNote{}
\end{adjustwidth}
\end{document}